\documentclass[aip,jmp,amsmath,amssymb,reprint]{revtex4-1}

\usepackage{graphicx}
\usepackage{booktabs}
\usepackage{here}

\usepackage{color}
 \makeatletter
\newcommand{\ccell}[3][]{%
  \kern-\fboxsep
  \if\relax\detokenize{#1}\relax
    \expandafter\@firstoftwo
  \else
    \expandafter\@secondoftwo
  \fi
  {\colorbox{#2}}%
  {\colorbox[#1]{#2}}%
  {#3}\kern-\fboxsep
}
\makeatother
\definecolor{cellgray}{gray}{0.9}

\begin{document}


\title[Accelerating Wave Function Convergence in Interactive Quantum Chemical Reactivity Studies]{Accelerating Wave Function Convergence in Interactive Quantum Chemical Reactivity Studies}

\author{Adrian H. M\"{u}hlbach}%
\affiliation{ 
ETH Z\"urich, Laboratorium f\"ur Physikalische Chemie, Vladimir-Prelog-Weg 2, 8093 Z\"urich, Switzerland
}%
\author{Alain C. Vaucher}
\affiliation{ 
ETH Z\"urich, Laboratorium f\"ur Physikalische Chemie, Vladimir-Prelog-Weg 2, 8093 Z\"urich, Switzerland
}%
\author{Markus Reiher}
 \email{markus.reiher@phys.chem.ethz.ch (corresponding author)}
\affiliation{ 
ETH Z\"urich, Laboratorium f\"ur Physikalische Chemie, Vladimir-Prelog-Weg 2, 8093 Z\"urich, Switzerland
}%

\date{7 December 2015}

\begin{abstract}
The inherently high computational cost of iterative self-consistent-field (SCF) methods 
proves to be a critical issue delaying visual and haptic feedback in real-time quantum chemistry. 
In this work, we introduce two schemes for SCF acceleration.
They provide a guess for the initial density matrix of the SCF procedure
generated by extrapolation techniques. SCF optimizations then converge in fewer iterations, which decreases the execution time 
of the SCF optimization procedure. 
To benchmark the proposed propagation schemes, we developed a test bed for performing quantum chemical calculations on sequences 
of molecular structures mimicking real-time quantum chemical explorations. 
Explorations of a set of six model reactions employing the semi-empirical methods PM6 and DFTB3 in this testing environment 
showed that the proposed propagation schemes achieved speedups of up to thirty percent as a consequence of a reduced number 
of SCF iterations. 
\end{abstract}

\maketitle

\setlength{\parindent}{0cm}
\setlength{\parskip}{0.6em plus0.2em minus0.1em}

\section{Introduction}

Over the last decades computer processing speed has reached a level where even personal computers are able to solve quantum chemical calculations on 
moderately sized molecules. Recently, we showed\cite{haag2013,haag2014a,haag2014b} that molecules with on the order of one hundred atoms can be calculated
on the millisecond timescale.
As a result, quantum chemical calculations on such molecular structures can be carried out and analyzed instantaneously.
In previous work, we implemented semi-empirical calculations in a real-time quantum chemistry framework.\cite{marti2009, haag2011, haag2013, haag2014a, haag2014b, vaucher2015a}
While such calculations are feasible with methods of density-functional theory,\cite{haag2013,luehr2015a} 
semi-empirical approaches that qualitatively reproduce all features of a Born--Oppenheimer potential energy surface
are an attractive target for real-time quantum chemistry.\cite{haag2014a}
In spite of their inherent approximations, the accuracy of these highly parametrized semi-empirical methods rivals that of 
standard quantum chemical methods.\cite{gaus2011, stewart2007, korth2011c, sedlak2013a, hostas2013a} 

In the real-time quantum chemistry framework\cite{haag2013}, chemical systems can be interactively explored 
with a proper hardware device such as a force-feedback (haptic) device (or an ordinary computer mouse)\cite{haag2014b, vaucher2015a}.
Results of the calculations such as the total electronic energy and the forces acting on the atoms are immediately transmitted back to the operator.
This allows for an immersive exploration of the potential energy surface.
Other interactive applications of quantum chemistry comprise the real-time optimization of molecular structures\cite{bosson2012} 
and interactive \textit{ab initio} molecular dynamics.\cite{luehr2015a}

Semi-empirical methods such as DFTB2,\cite{elstner1998} DFTB3,\cite{gaus2011} and PM6\cite{stewart2007}, which are available in our 
real-time quantum chemistry framework\cite{vaucher2015a}, implement prototypical orbital models that require iterative self-consistent-field (SCF) 
optimizations.
SCF methods are not trivial to apply in real time due to their iterative nature, which results in unpredictable calculation times.
Another problem is the difficulty of convergence control for molecular structures that are far away from their equilibrium geometry.
As these issues do not vanish by application of convergence acceleration techniques, the SCF procedure benefits from an initial 
guess density matrix close to the converged density matrix. 

In this work, we develop efficient density matrix propagation schemes that can reduce the number of SCF iteration steps by 
extrapolation of converged density matrices obtained for preceding molecular structures of a real-time exploration.
This work is organized as follows.
In section \ref{sct:rtqc}, we review the difficulties of the application of SCF methods within the real-time quantum chemistry framework.
Then, in section \ref{sct:propagation}, we discuss different possibilities for their acceleration and introduce two schemes 
to provide initial density matrices for the SCF optimization procedure.
In section \ref{sct:results}, we then study their application on a set of model reactions.

\section{Self-consistent field iterations in real-time quantum chemistry}
\label{sct:rtqc}

The real-time quantum chemistry framework\cite{haag2013} is an immersive tool for the interactive exploration of chemical reactivity. 
As such, it contains a component for quantum chemical calculations combined with one or several components for the immersive 
interaction with the molecular system under consideration.

\begin{figure}[htb]
\centering
\includegraphics[width=7cm,keepaspectratio=true]{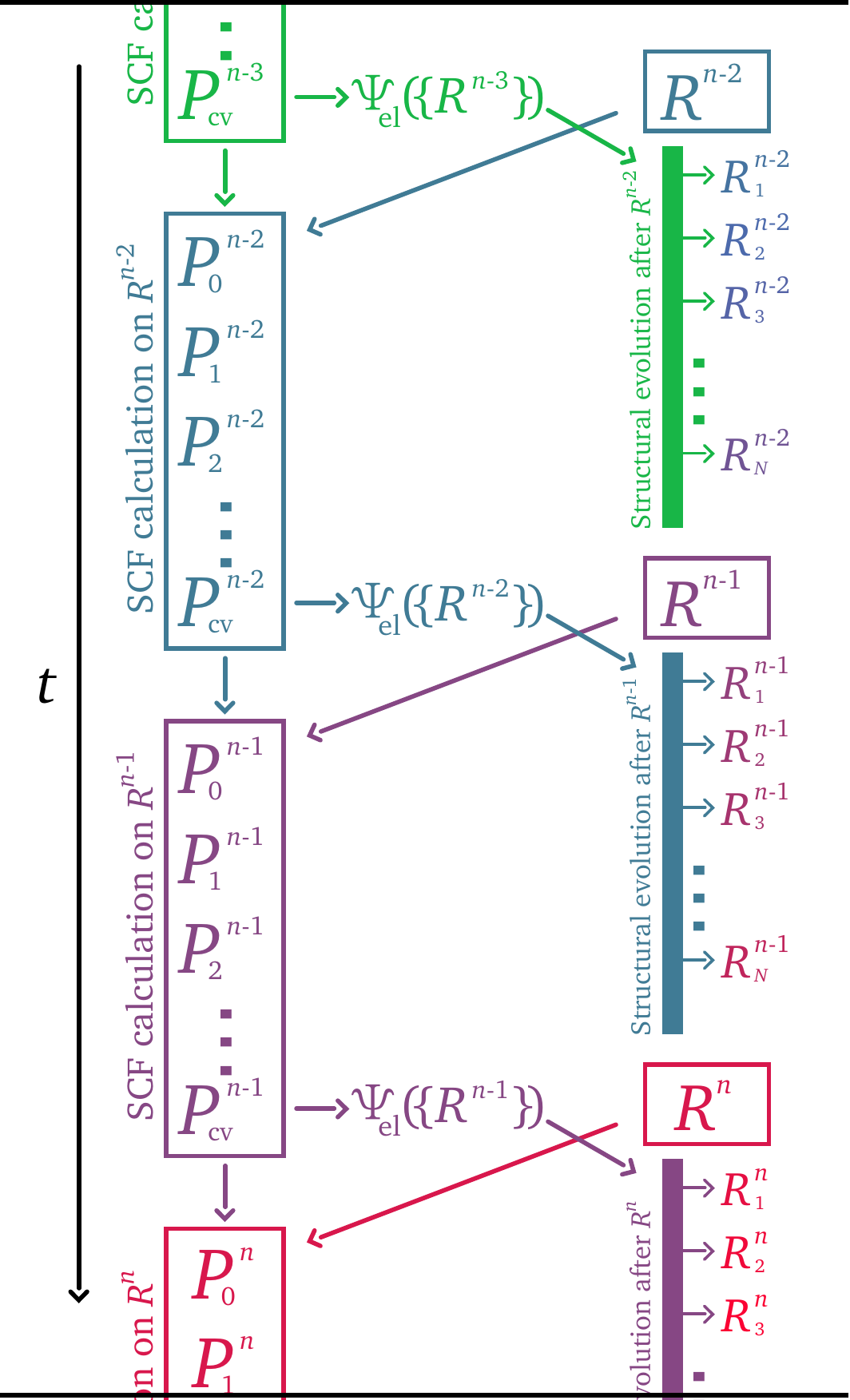}
\caption{Handling of SCF calculations in the real-time quantum chemistry framework with increasing time $t$.
         On the left, each box indicates an SCF calculation that iteratively optimizes the density matrix ($\boldsymbol{P}_i^j$ in the
         $i$-th iteration step for the $j$-th structure) until it is converged ($\boldsymbol{P}^j_\mathrm{cv}$).
	 Each SCF calculation yields orbitals from which the electronic wave function $\Psi_\mathrm{el} ( \{ \boldsymbol{R} \} )$ and thus quantum chemical properties are calculated (middle).
	 On the right, each $\boldsymbol{R}_i^j$ symbol represents a molecular structure that is generated for visual representation. 
	 SCF calculations are executed for the ones surrounded by a box. 
         It can be seen that there is a delay between the start of the calculation on a structure and the delivery of the quantum chemical properties for this structure.
 }
\label{scf_haptic_scheme}
\end{figure}

An operator can create a series of molecular structures through interactive structure manipulation.
There are two aspects of such an emergent structural evolution: (i) the structural relaxation by structure optimization and 
(ii) the structural perturbation introduced by the operator with input devices such as a computer mouse or a haptic device.
In our current implementation,\cite{vaucher2015a} relaxation is handled by a steepest-descent algorithm where the nuclei are moved 
along their negative gradient.
The quantum chemical calculations run continuously in the background for consecutive structures visited during a reactivity exploration. 
When applying SCF methods, the execution time of a single-point calculation is unpredictable because of the iterative optimization of the 
electronic structure for a fixed molecular structure. This means that a self-consistent density matrix $\boldsymbol{P}$
calculated from the molecular orbital coefficients is obtained.
Hence, it is necessary to decouple the quantum chemical calculations from the structure evolution in order to preserve the 
immersion in the reactivity exploration process.\cite{vaucher2015a} 
As a consequence, feedback (for a given molecular structure) in the form of structural evolution and haptic force rendering is based on the electronic structure of an earlier molecular structure.
An illustration of the procedure is shown in Fig.~\ref{scf_haptic_scheme}.

In a recent work\cite{vaucher2015a}, we introduced a strategy to ensure reliable reactivity exploration when feedback is provided with an unpredictable time delay.
In the present work, we introduce schemes to allow for more frequent feedback by reducing the execution time of electronic structure optimizations.
To achieve this, we here develop techniques that aim at improved guesses for the initial density matrix $\boldsymbol{P}_0$ 
to accurately approximate the converged matrix
$\boldsymbol{P}_\mathrm{cv}$ so that the number of SCF iteration steps is reduced.

\section{The LS-R and the LS-S density matrix propagation schemes}
\label{sct:propagation}

Acceleration of SCF convergence has been the objective of research efforts for many years.
In addition to reducing the number of SCF iterations required for convergence, convergence-acceleration schemes 
are essential to reach self-consistency at all.

One option to accelerate SCF convergence is to exploit, for a molecular structure under consideration, the information produced by successive iterations of the SCF procedure to allow for better guesses for the subsequent iterations.
This option is, for example, realized in the level-shifting method\cite{saunders1973}, in the direct inversion of the iterative subspace (DIIS) algorithm\cite{pulay1980,pulay1982}, in energy-DIIS (EDIIS)\cite{kudin2002}, in augmented-DIIS (ADIIS)\cite{hu2010}, in the augmented Roothaan--Hall (ARH) method\cite{host2008} and in linear-expansion shooting techniques (LIST)\cite{wang2011,chen2011a}.
Another approach is the orbital transformation method\cite{vandevondele2003}, 
which performs transformations of the molecular orbitals that avoid the diagonalization of the Fock matrix.

When SCF calculations must be performed for several similar structures, another option for SCF acceleration is to exploit the converged information of related molecular structures.
This is particularly helpful when the electronic structure of several consecutive molecular structures needs to be calculated, as for example in \textit{ab initio} molecular dynamics, geometry optimization and, in our case, interactive reactivity explorations.
In such cases, two different approaches to SCF acceleration are prevalent.
On the one hand, optimization-free calculations have been in the focus of research, particularly in the field of molecular dynamics.
This approach is for example applied in Car--Parrinello molecular dynamics\cite{car1985}, in density matrix dynamics\cite{schlegel2001,iyengar2001,schlegel2002a} and in approaches related to extended-Lagrangian Born--Oppenheimer molecular dynamics\cite{niklasson2012,souvatzis2014a}.
On the other hand, several strategies aim at a reduction of the number of SCF iterations by providing guesses for the first SCF iteration relying on the optimized electronic structures of previous molecular structures.
For example, Pulay and Fogarasi extrapolate the Fock matrix\cite{pulay2004}, while Atsumi and Nakai\cite{atsumi2008, atsumi2010} extraploate the molecular orbitals, and VandeVondele et al.\ extrapolate the density matrix in combination with the orbital transformation method mentioned above.\cite{vandevondele2005}

In our real-time framework, the last converged density matrix was injected as a density matrix guess for the new structure and then 
optimized by the DIIS algorithm in the SCF optimization procedure.
For further acceleration, many of the approaches mentioned above are not applicable in real-time quantum chemistry applications.
In fact, most of them rely on successive structures produced in short time or distortion steps, which are not realized in real-time quantum chemistry
as the step sizes are defined by the action of the operator.
Due to its special needs, real-time quantum chemistry requires schemes tailored for this special type of configuration-space exploration.
Unlike in molecular dynamics, time reversibility is not a concern in real-time quantum chemistry because excess energy can always be removed 
by structure optimization. Most importantly, the operator can create structural changes that do not correspond to small coordinate distortions.
In the following two sections, we propose two schemes that extrapolate converged density matrices to provide an improved guess for the initial density matrix of the SCF procedure.

\subsection{Least-squares propagation with nuclear coordinates}
The scheme we propose in this section is inspired by the least-squared prediction of the molecular orbitals (LSMO) technique developed by Atsumi and Nakai\cite{atsumi2010}. 
The main idea behind their method is to construct the new guess molecular orbitals as a linear combination from the converged molecular orbitals of 
previous (structure) steps.
The molecular orbital coefficients are obtained by least-squares extrapolation in a procedure analogous to DIIS with an 
error vector derived from differences in the coordinates of the molecular structures considered. 
To account for the crossing and mixing of the molecular orbitals, Atsumi and Nakai multiply the molecular
orbital coefficients matrices with auxiliary transform matrices in the linear combination.

In the present work, we introduce a modification of LSMO where the density matrix is propagated instead of the molecular orbitals.
The density matrix carries sufficient information for starting the SCF procedure and avoids the need and computational cost of auxiliary transform matrices.
The ($n$+1)-th molecular structure with coordinates $\boldsymbol{R}^{n+1}$, for which a density matrix 
guess $\boldsymbol{P}_0^{n+1}$ is to be produced, can be described as the sum of the linear combination of $K$ previous 
structures and a residual error $\boldsymbol{E}$:
      \begin{equation}
      \boldsymbol{R}^{n+1} = \boldsymbol{E} + \sum_{k = 0}^{K-1} c_{k} \boldsymbol{R}^{n-k}.
      \end{equation}
The coefficients $c_{k}$ are obtained by minimizing the residual error $\boldsymbol{E}$ with a least-squares minimization, as in DIIS:
      \begin{equation}
      \begin{pmatrix}
	0 & -1 & -1 & \cdots & -1\\
	-1 & B_{00} & B_{01} & \cdots & B_{0K}\\
	-1 & B_{10} & B_{11} & \cdots & B_{1K}\\
	\vdots & \vdots & \vdots &  \ddots & \vdots\\
	-1 & B_{K0} & B_{K1} & \cdots & B_{KK}
      \end{pmatrix}
      \begin{pmatrix}
	-\lambda\\
	c_{0}\\
	c_{1}\\
	\vdots\\
	c_{K}\\
      \end{pmatrix}
      = 
      \begin{pmatrix}
	-1\\
	0\\
	0\\
	\vdots\\
	0\\
      \end{pmatrix}.
      \label{test}
      \end{equation}
Here, $\lambda$ denotes a Lagrange multiplicator and $B_{ij}$ is given by
\begin{equation}
\begin{aligned}
	B_{ij} = B_{ji} &= \langle \boldsymbol{R}^{n-i} - \boldsymbol{R}^{n+1} \mid \boldsymbol{R}^{n-j} - \boldsymbol{R}^{n+1} \rangle \\
	&= \sum_{k=1}^{3N} (R_k^{n-i} - R_k^{n+1})(R_k^{n-j} - R_k^{n+1}).
\end{aligned}
\end{equation}
where $N$ denotes the number of atoms.
The new guess density matrix is then calculated from the linear combination of the $K$ previously converged density 
matrices $\boldsymbol{P}_\mathrm{cv}^{n-k}$:
\begin{equation}
      \boldsymbol{P}_0^{n+1} = \sum_{k = 0}^{K-1} c_{k} \boldsymbol{P}_\mathrm{cv}^{n-k}.
\end{equation}

Partly restoring idempotency of the guess density matrix with the McWeeny purification algorithm\cite{mcweeny1959, mcweeny1960} turned out to be beneficial.
In an orthonormal basis, the McWeeny purification algorithm reads
\begin{equation}
\boldsymbol{P}' = 3 \boldsymbol{P}^2 - 2 \boldsymbol{P}^3
\end{equation}
and, in a non-orthonormal basis, 
\begin{equation}
\boldsymbol{P}' = 3 \boldsymbol{PSP} - 2 \boldsymbol{PSPSP},
\end{equation}
where $\boldsymbol{S}$ denotes the overlap matrix.
In our propagation scheme, the McWeeny purification algorithm can be applied a variable number of times.
This propagation scheme will be abbreviated as LS-R$^{\gamma}_{K}$ (\underline{l}east-\underline{s}quares propagation with \textbf{\underline{R}}) with $\gamma$ denoting the number of McWeeny purifications applied and
$K$ being the number of preceding structures and converged density matrices considered for the linear extrapolation.

\subsection{Least-squares propagation with overlap matrices}
The second scheme we propose is very similar to the LS-R scheme presented above.
Instead of employing changes in the nuclear coordinates for determining the coefficients of the linear combination, this information is obtained from the overlap matrices.
The matrix entries $B_{ij}$ then reads
\begin{equation}
\begin{aligned}
	B_{ij} = B_{ji} &= \langle \boldsymbol{S}^{n-i} - \boldsymbol{S}^{n+1} \mid \boldsymbol{S}^{n-j} - \boldsymbol{S}^{n+1} \rangle \\
      &= \sum_{\mu\nu} (S_{\mu\nu}^{n-i} - S_{\mu\nu}^{n+1})(S_{\mu\nu}^{n-j} - S_{\mu\nu}^{n+1}).
\end{aligned}
\end{equation}
This is the only modification to the LS-R propagation scheme.
We abbreviate this second propagation scheme as LS-S$^{\gamma}_{K}$ (\underline{l}east-\underline{s}quares propagation 
with \textbf{\underline{S}}) with $\gamma$ denoting the number of McWeeny purification steps as before.
In analogy to LS-R, $K$ denotes the number of previous overlap matrices and converged density matrices considered for the linear extrapolation.

\section{Computational methodology}
\label{sct:results}

\subsection{The testing environment}\label{testenv}
We developed a testing environment in order to compare the propagation schemes in a reproducible manner.
This C++ program simulates a real-time reactivity exploration by generating sequences of structures for model 
reactions and performing quantum chemical calculations for structures along these extrapolation coordinates.

For a given model reaction, one first defines one or several atom pairs whose internuclear distance will be 
changed to construct some path in configuration space from reactant structures to products. 
Then, the sequence of structures is constructed from the starting structure $\boldsymbol{R}^{0}$ in a recursive manner. 
The structure $\boldsymbol{R}^{n+1}$ is obtained from the structure $\boldsymbol{R}^{n}$ as follows. 
First, the internuclear distances of the previously defined atom pairs are modified according to a specified step size. 
Then, this structure is subjected to several steps of a (constraint) steepest descent structure optimization utilizing 
the gradient of the energy with respect to nuclear coordinates. 
During this constraint structure relaxation the selected-pair(s) internuclear distances remain fixed.
This reaction trajectory is then used to analyze the performance of the propagation schemes.

Each propagation step is followed by a quantum chemical calculation of the molecular structure relying on the PM6 and DFTB3 methods, which were recently implemented in our group as C++ libraries.\cite{vaucher2015a}
Although these methods are semi-empirical, similar results are expected for first-principles SCF methods such as Kohn--Sham density functional theory, as they share the same formalism.
For some reactions, DFTB3 could not be employed because of the lack of adequate parameters.

In addition to the LS-R and LS-S schemes, the SCF convergence was accelerated by the DIIS algorithm.\cite{pulay1980,pulay1982}
For a calculation to be considered converged, the Frobenius norm of the difference between two contiguous density matrices $\boldsymbol{P}_{i-1}$ and $\boldsymbol{P}_i$ had to be below $10^{-5} M^2$, with $M$ being the total number of atomic orbitals (basis functions); 
i.e.\ the following condition must be fulfilled:
\begin{align}
	\frac{1}{M^2} \sqrt{\sum_\mu^M \sum_\nu^M \left( P_{\mu\nu,i} - P_{\mu\nu,i-1} \right) ^2} < 10^{-5}.
\end{align}

\subsection{Model reactions}\label{model_reactions}

\begin{figure}[htb]
\centering
\includegraphics[width=.48\textwidth]{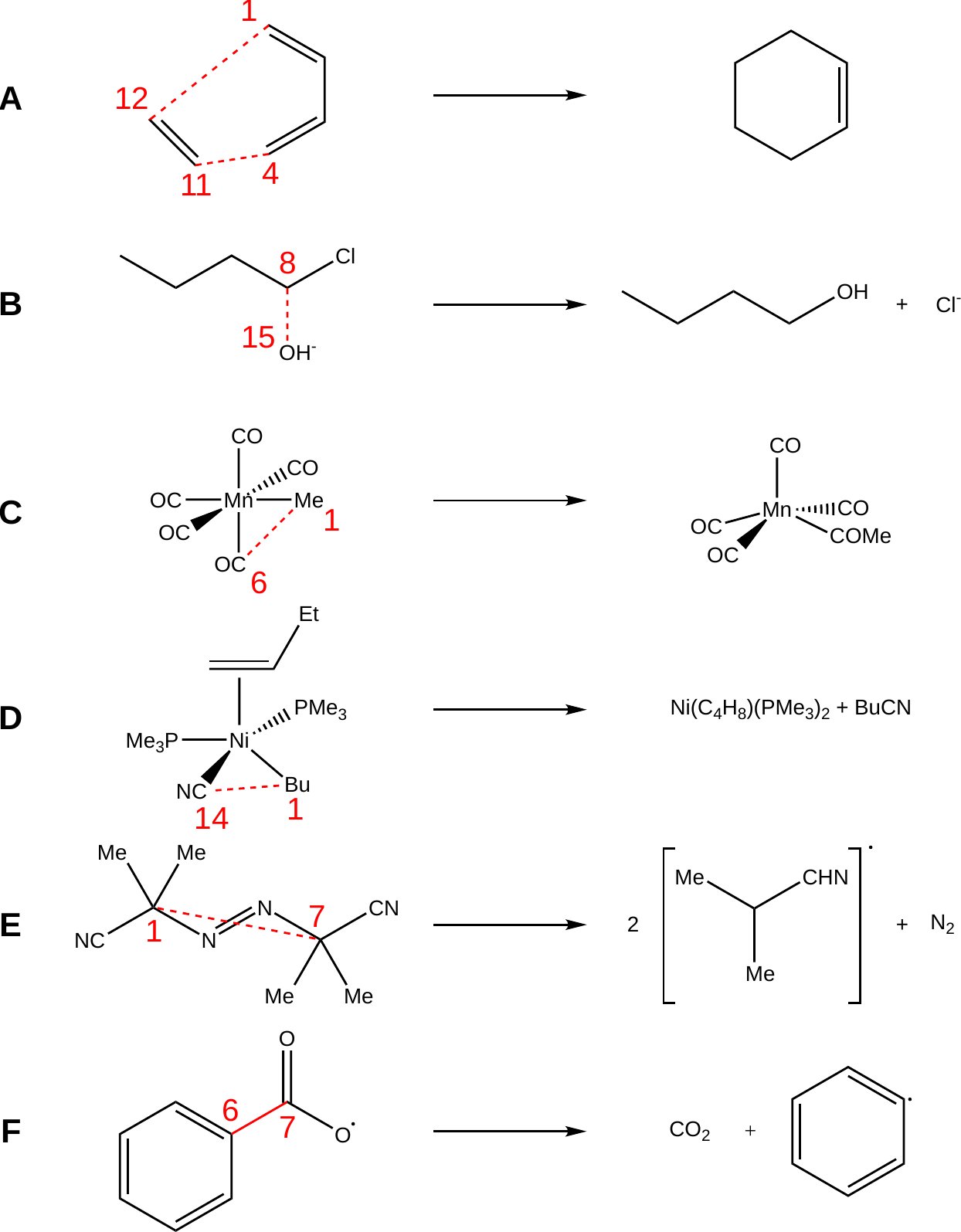}
\caption{Model reactions studied in this work: a Diels--Alder reaction (\textbf{A}), a nucleophilic substitution (\textbf{B}), a migratory insertion (\textbf{C}), a reductive elimination (\textbf{D}), the decomposition of AIBN (\textbf{E}) and the elimination of CO$_2$ in the benzoyl radical (\textbf{F}).
The atom pairs whose internuclear distance were scanned are marked in red.}
\label{reactions}
\end{figure}

To benchmark the propagation schemes in the testing environment, we chose six model reactions.
They are shown in Fig.\ \ref{reactions}.
The Cartesian coordinates for all the sequences of structures can be found in the Supporting Information.
The reactions \textbf{A}, \textbf{B}, \textbf{C} and \textbf{D} were treated in the spin-restricted formalism and the reactions \textbf{E} and \textbf{F} in the spin-unrestricted formalism.

\paragraph*{A: Diels--Alder reaction}
As an example of a concerted pericyclic reaction we chose the Diels--Alder reaction of ethene and butadiene (\textbf{A} in Fig.\ \ref{reactions}).
The step sizes for the scans between the atom pairs C$^{1}$--C$^{12}$ and C$^{4}$--C$^{11}$ were 0.1 \AA{} and 0.06 \AA{}, respectively.
Cyclohexene was obtained as the product.

\paragraph*{B: Nucleophilic substitution S$_\text{N}$2}
For the nucleophilic substitution reaction we chose a nucleophilic attack of a hydroxide anion onto 1-chlorbutane with subsequent ejection of an chloride anion (\textbf{B} in Fig.\ \ref{reactions}).
We performed a linear scan on the atom pair C$^{8}$--O$^{15}$ with a step size of 0.1 \AA{}.

\paragraph*{C: Insertion of carbon monoxide into a metal-alkyl bond}
For the migratory insertion reaction we chose the Mn(CO)$_{5}$Me complex.
One of the carbon monoxide ligands is inserted into the manganese-methyl bond (\textbf{C} in Fig.\ \ref{reactions}).
A linear scan on the atom pair C$^{1}$--O$^{6}$ with a step size of 0.1 \AA{} was performed.

\paragraph*{D: Carbon-carbon bond formation through reductive elimination}
A model for a reductive elimination reaction is the NiBu(C$_{4}$H$_{8}$)(PMe$_{3}$)$_{2}$CN complex.
The cyanide and the butyl ligand form 1-cyanobutane via reductive elimination (\textbf{D} in Fig.\ \ref{reactions}).
A linear scan was performed on the atom pair C$^{1}$--C$^{14}$ with a step size of 0.1 \AA{}.

\paragraph*{E: Decomposition of AIBN}
As an example for a radical reaction, we chose the decomposition of azobisisobutyronitrile (AIBN) (\textbf{E} in Fig.\ \ref{reactions}).
Starting with AIBN, we obtain N$_2$ and two 2-cyanoprop-2-yl radicals after decomposition.
The step size for the scan of the atom pair C$^{1}$--C$^{7}$ was 0.1 \AA{}.

\paragraph*{F: Elimination of carbon dioxide in the benzoyl radical}
Another radical reaction is the elimination of CO$_2$ from the benzoyl radical to generate the phenyl radical.
A linear scan was performed on the atom pair C$^{6}$--C$^{7}$ with a step size of 0.1 \AA{}.

\subsection{Efficiency and accuracy measures}
\label{sct:dr}

To compare the different schemes, we define five measures. 
They can be evaluated for each single propagation step that delivers a guess density matrix for the molecular structure $n$.
Accordingly, they carry a subscript $n$ that indicates the corresponding propagation step.

\paragraph*{Execution time:}
The execution time $t_n$ for step $n$ is the sum of the time of the propagation of the density matrix, $t_{\mathrm{prop},n}$, and the time needed for the subsequent calculation to reach convergence, $t_{\mathrm{SCF},n}$:
\begin{equation}
	t_n = t_{\mathrm{prop}, n} + t_{\mathrm{SCF}, n}
\end{equation}

\paragraph*{Number of iterations:}
The number of iterations $n_{\mathrm{it},n}$ is the number of SCF iterations required to reach convergence for the 
orbitals of molecular structure $n$.

The energy and gradient obtained without SCF optimization by evaluating the guess density matrix at the new molecular structure were also monitored because they are good indicators of how close the guess is to the converged density matrix.

\paragraph*{Accuracy of predicted energies:}
To assess the energy accuracy of a propagation step, the indicator $(\Delta E)_n$ evaluates the difference of predicted and converged energies, 
$E_{\mathrm{pred},n}$ and $E_{\mathrm{cv},n}$, respectively:
\begin{equation}
	(\Delta E)_n = |E_{\mathrm{pred}, n} - E_{\mathrm{cv}, n}|
\end{equation}

\paragraph*{Accuracy of predicted gradients:}
To measure the accuracy of the predicted gradients we introduce two separate measures quantifying the amplitude and the angular errors of the predicted gradient vectors.
The error angle $\Phi_n$ is calculated as the angle between the predicted and the converged gradient vectors for each of the $N$ atoms:
\begin{equation}
	\Phi_n = N^{-1} \sum^{N}_{j=1}{\arccos\left(\frac{\langle{\nabla}_j E_{\mathrm{pred}, n}| {\nabla}_j E_{\mathrm{cv}, n}\rangle} {|{\nabla}_j E_{\mathrm{pred}, n}| \, |{\nabla}_j E_{\mathrm{cv}, n}|}\right)}
\end{equation}
The amplitude error $\Theta_n$ is calculated as the vector norm deviation of the predicted gradient vector from the converged gradient vector for each of the $N$ atoms:
\begin{equation}
	\Theta_n = N^{-1} \sum^{N}_{j=1}{\left|\frac{|{\nabla}_j E_{\mathrm{pred}, n}|}{|{\nabla}_j E_{\mathrm{cv}, n}|} - 1 \right|}
\label{amplitudes}
\end{equation}
This amplitude error is given as a percental deviation.

The measures for the reaction as a whole are given as arithmetic means of the measures for each of the $N_P$ steps:
\begin{align}
	\label{dr:speed}
	t &= N_P^{-1} \sum^{N_P}_{n=1} t_n, \\
	\label{dr:iterations}
	n_\mathrm{it} &= N_P^{-1} \sum^{N_P}_{n=1} n_{\mathrm{it},n}, \\
	\label{dr:energy}
	\Delta E &= N_P^{-1} \sum^{N_P}_{n=1} (\Delta E)_n, \\
	\label{dr:angle}
	\Phi &= N_P^{-1} \sum^{N_P}_{n=1}{\Phi_n}, \\
	\label{dr:amplitude}
	\Theta &= N_P^{-1} \sum^{N_P}_{n=1}{\Theta_n}.
\end{align}

\section{Results}

\subsection{Optimization of propagation parameters}\label{optimization}
The LS-R and LS-S schemes each have two parameters that need to be optimized for ideal performance: 
the number of preceding structures to consider and the number of McWeeny purifications applied.
The Supporting Information contains tables that compare the results obtained with the parameter combinations tested during the parameter optimization.

\subsubsection*{LS-R propagation}\label{results_lsr}
The results of the LS-R propagation are presented in Tables S1--S5 of the Supporting Information.
The highest SCF accelerations are observed for LS-R$^1_4$ and LS-R$^1_5$.
For the studied model reactions, these two propagation schemes are capable of reducing the number of SCF iterations by up to 50\% and they achieve speedups of approximately 30\%.
The repeated application of the McWeeny purification results in increasingly better predicted energies and gradients.

\subsubsection*{LS-S propagation}\label{results_lss}
The results of the LS-S propagated model reactions are presented in Tables S6--S10 of the Supporting Information.
The highest SCF accelerations are observed for LS-S$^1_4$ and LS-S$^1_5$.
They also reduce the number of SCF iterations by up to 50\% and achieve speedups of approximately 30\%.
Also, the repeated application of the McWeeny purification results in increasingly better predicted energies and gradients.

\subsection{Comparison of the optimized propagation schemes}

When comparing the four propagation schemes LS-R$_4^1$, LS-R$_5^1$, LS-S$_4^1$ and LS-S$_5^1$ with respect to the number of SCF iterations (Table~\ref{comp-it}) and the computational times (Table~\ref{comp-t}) they all provide considerable improvements compared to 
simply taking the last converged density matrix in combination with the DIIS algorithm (denoted by the symbol ``\o{}'' in the tables).
Within our testing environment these four propagation schemes perform equally well.

\begin{table}[H]
      \caption{Number of iterations $n_\mathrm{it}$ of Eq.\ (\ref{dr:iterations}) for the propagation schemes with optimal parameters.
               The values are averages over the different molecular structures along the reaction coordinates.}
      \label{comp-it}
      \centering\small
      \begin{tabular}{cccccccccccc} 
      \hline\hline
      \multicolumn{1}{c}{$n_\mathrm{it}$} & &  \multicolumn{6}{c}{PM6} &  & \multicolumn{3}{c}{DFTB3}\\
      \hline
      & &  \textbf{A} & \textbf{B} & \textbf{C} & \textbf{D} & \textbf{E} & \textbf{F} & & \textbf{A} & \textbf{E} & \textbf{F}\\ 
      \hline\hline
      LS-R$_4^1$ & &  3.3 & 3.4 & 4.6 & 6.2 & 3.2 & 5.8 &  & 3.3 & 4.8 & 3.3 \\
      LS-R$_5^1$ & &  3.6 & 3.6 & 4.2 & 5.9 & 3.4 & 7.1 &  & 3.3 & 4.8 & 3.6 \\
      \hline
      LS-S$_4^1$ & &  3.3 & 3.2 & 4.5 & 5.9 & 2.7 & 4.8 &  & 3.4 & 4.7 & 3.2 \\
      LS-S$_5^1$ & &  3.5 & 3.5 & 4.2 & 6.1 & 3.6 & 5.8 &  & 3.3 & 4.5 & 3.1 \\
      \hline
      \o & &  6.2 & 6.4 & 7.8 & 9.2 & 7.6 & 9.8 &  & 4.8 & 5.9 & 4.5 \\
      \hline\hline
      \end{tabular}
\end{table}

\begin{figure}[htbp]
\centering
\includegraphics[width=0.48\textwidth]{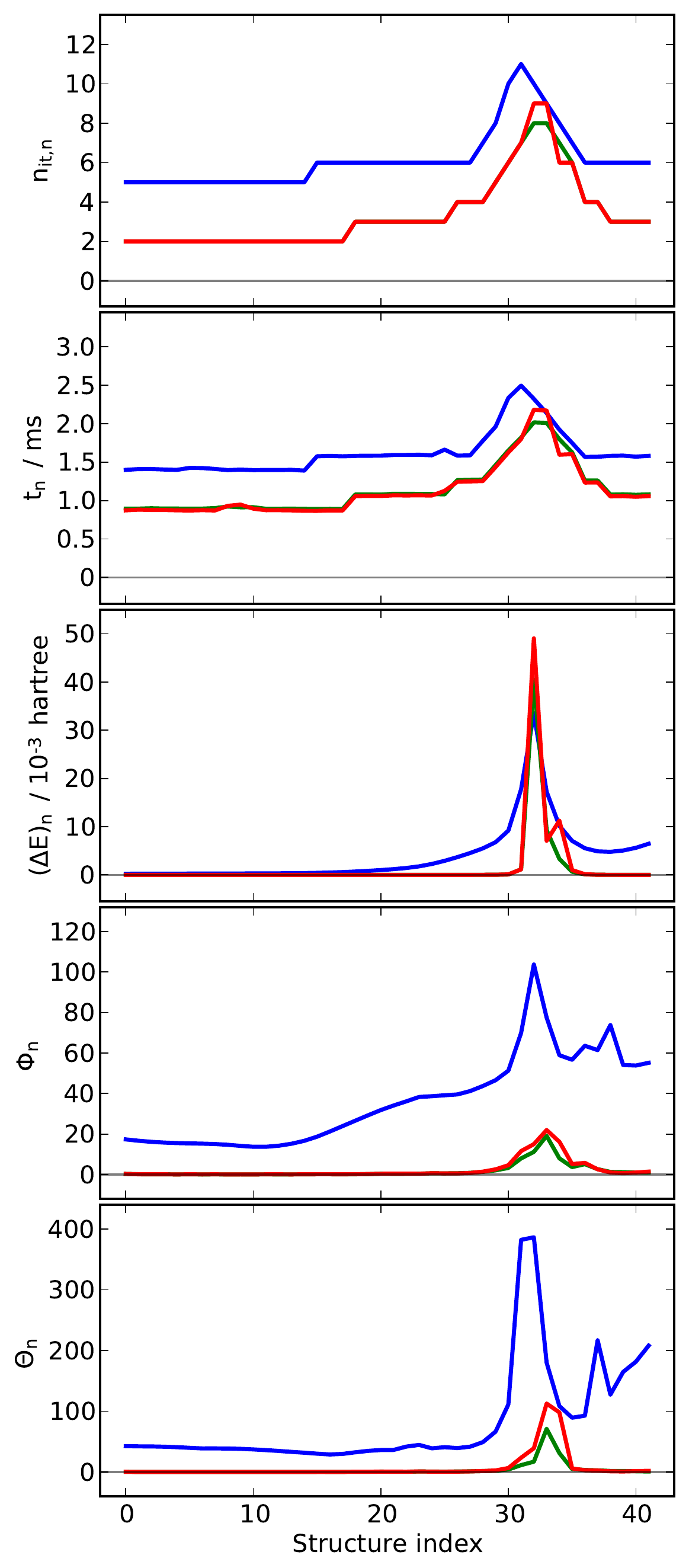}
\caption{From top to bottom: Numbers of iterations $n_{\mathrm{it},n}$, execution times $t_n$, errors in the predicted energy $(\Delta E)_n$, in the gradient angle $\Phi_n$ and in the gradient amplitude $\Theta_n$ for propagations with optimized parameters over the course of the Diels--Alder reaction (\textbf{A} in Fig.\ \ref{reactions}) calculated with PM6.
      The abscissa displays the reaction coordinate characterized by the indices of the structures encountered in the reaction.
      The blue curves are the results from the calculations in which no propagation was applied.
      The green and red curves represent the LS-S$^1_4$ and LS-R$^1_4$, respectively.
      In all cases, the results are less accurate at the transition state of the reaction (approximately at the structure with index 32).
      }
\label{comparison}
\end{figure}

Fig.\ \ref{comparison} illustrates the efficiency of the single propagation steps for the Diels--Alder reaction, corresponding to the visited molecular structures along the reaction coordinate, with PM6.
It shows that the propagation schemes LS-R$_4^1$ and LS-S$_4^1$ accelerate the SCF calculations for all structures visited. 
The acceleration is less pronounced for structures near the transition state because of the rapidly changing electronic structure for consecutive steps at the transition state.

\begin{table}[H]
      \caption{Computational time $t$ (in ms) Eq.\ (\ref{dr:speed}) for the propagation schemes with optimal parameters.}
      \label{comp-t}
      \centering\small
      \begin{tabular}{cccccccccccc} 
      \hline\hline
      \multicolumn{1}{c}{$t$} & &  \multicolumn{6}{c}{PM6} & &  \multicolumn{3}{c}{DFTB3}\\
      \hline
      & &  \textbf{A} & \textbf{B} & \textbf{C} & \textbf{D} & \textbf{E} & \textbf{F} & &  \textbf{A} & \textbf{E} & \textbf{F}\\ 
      \hline
      LS-R$_4^1$ & &  1.13 & 1.81 & 4.59 & 38.42 & 3.68 & 4.02 & &  0.92 & 8.67 & 2.66 \\
      LS-R$_5^1$ & &  1.17 & 1.89 & 4.39 & 37.41 & 3.78 & 4.34 & &  0.92 & 8.69 & 2.89 \\
      \hline
      LS-S$_4^1$ & &  1.14 & 1.80 & 4.58 & 37.71 & 3.41 & 3.81 & &  0.98 & 8.57 & 2.63 \\
      LS-S$_5^1$ & &  1.20 & 1.90 & 4.41 & 38.45 & 4.06 & 4.27 & &  0.97 & 8.26 & 2.55 \\
      \hline
      \o & &  1.62 & 2.70 & 6.50 & 50.17 & 6.46 & 6.25 & &  1.19 & 10.16 & 3.35 \\
      \hline\hline
      \end{tabular}
\end{table}

The four propagation schemes LS-R$_4^1$, LS-R$_5^1$, LS-S$_4^1$ and LS-S$_5^1$ also deliver better predicted energies and gradients compared to the application of the DIIS algorithm alone (Tables~\ref{comp-e}--\ref{comp-amp}). 

\begin{table}[H]
      \caption{Error in the energy $\Delta E$ (in $10^{-5}$ Hartree) of Eq.\ (\ref{dr:energy}) for the propagation schemes with optimal parameters.}
      \label{comp-e}
      \centering\small
      \begin{tabular}{cccccccccccc} 
      \hline\hline
      \multicolumn{1}{c}{$\Delta E$} & & \multicolumn{6}{c}{PM6} & & \multicolumn{3}{c}{DFTB3}\\
      \hline
      & & \textbf{A} & \textbf{B} & \textbf{C} & \textbf{D} & \textbf{E} & \textbf{F} & & \textbf{A} & \textbf{E} & \textbf{F}\\ 
      \hline
      LS-R$_4^1$ & & 166 & 16.2 & 9.9 & 913 & 1.1 & 4.0 & & 3.1 & 8.5 & 1.4 \\
      LS-R$_5^1$ & & 202 & 15.6 & 7.3 & 904 & 1.0 & 5.2 & & 1.9 & 6.6 & 1.4 \\
      \hline
      LS-S$_4^1$ & & 131 & 7.2 & 9.6 & 571 & 1.0 & 0.9 & & 2.9 & 2.8 & 0.9 \\
      LS-S$_5^1$ & & 133 & 4.7 & 6.5 & 512 & 1.0 & 3.1 & & 1.5 & 2.4 & 0.9 \\
      \hline
      \o & & 394 & 499 & 801 & 1117 & 314 & 88.8 & & 933 & 348 & 37.6 \\
      \hline\hline
      \end{tabular}
\end{table}

\begin{table}[H]
      \caption{Gradient angle error $\Phi$ (in degrees) of Eq.\ (\ref{dr:angle}) for the propagation schemes with optimal parameters.}
      \label{comp-ang}
      \centering\small
      \begin{tabular}{cccccccccccc} 
      \hline\hline
      \multicolumn{1}{c}{$\Phi$} & & \multicolumn{6}{c}{PM6} & & \multicolumn{3}{c}{DFTB3}\\
      \hline
      & & \textbf{A} & \textbf{B} & \textbf{C} & \textbf{D} & \textbf{E} & \textbf{F} & & \textbf{A} & \textbf{E} & \textbf{F}\\ 
      \hline
      LS-R$_4^1$ & & 2.3 & 2.4 & 6.4 & 8.9 & 1.4 & 4.5 & & 1.2 & 8.0 & 1.4 \\
      LS-R$_5^1$ & & 2.0 & 2.1 & 3.7 & 7.9 & 0.9 & 4.8 & & 1.1 & 7.6 & 1.3 \\
      \hline
      LS-S$_4^1$ & & 1.8 & 2.2 & 6.6 & 8.4 & 1.3 & 0.8 & & 1.1 & 6.7 & 0.7 \\
      LS-S$_5^1$ & & 1.5 & 1.7 & 3.8 & 7.3 & 1.3 & 3.1 & & 1.0 & 7.1 & 1.1 \\
      \hline
      \o & & 36.2 & 60.5 & 86.6 & 51.3 & 52.9 & 6.1 & & 23.0 & 48.1 & 5.4 \\
      \hline\hline
      \end{tabular}
\end{table}

\begin{table}[H]
      \caption{Gradient amplitude error $\Theta$ (in percent) of Eq.\ (\ref{dr:amplitude}) for the propagation schemes with optimal parameters.}
      \label{comp-amp}
      \centering\small
      \begin{tabular}{cccccccccccc} 
      \hline\hline
      \multicolumn{1}{c}{$\Theta$} & & \multicolumn{6}{c}{PM6} & & \multicolumn{3}{c}{DFTB3}\\
      \hline
      & & \textbf{A} & \textbf{B} & \textbf{C} & \textbf{D} & \textbf{E} & \textbf{F} & & \textbf{A} & \textbf{E} & \textbf{F}\\ 
      \hline
      LS-R$_4^1$ & & 7.3 & 3.3 & 10.4 & 21.4 & 2.2 & 5.5 & & 1.4 & 21.6 & 2.0 \\
      LS-R$_5^1$ & & 4.9 & 2.7 & 5.9 & 19.4 & 2.1 & 7.5 & & 1.2 & 17.7 & 4.8 \\
      \hline
      LS-S$_4^1$ & & 3.9 & 2.9 & 10.3 & 17.7 & 1.5 & 1.2 & & 1.3 & 13.0 & 1.4 \\
      LS-S$_5^1$ & & 2.5 & 2.3 & 5.6 & 15.6 & 2.8 & 3.3 & & 1.2 & 11.6 & 1.0 \\
      \hline
      \o & & 81.3 & 111 & 548 & 172 & 663 & 47.2 & & 45.0 & 430 & 344 \\
      \hline\hline
      \end{tabular}
\end{table}

\subsection{Step size stability}\label{results_steps}
So far, the results presented in this section all originate from calculations on reactions where atomic distances were changed by approximately 0.1 \AA{} from one calculation to the other (see section \ref{model_reactions}).
This step size is comparable to the structural changes occurring during a standard real-time reactivity exploration.
We also assessed the performance of the previously selected propagation schemes for larger step sizes.
In the real-time quantum chemistry framework, this corresponds to larger structural changes between consecutive calculations, 
which can be caused by faster manipulations of the operator.
The results are given in Tables~S11-S15 of the Supporting Information.
With double step size, the LS-R$^1_4$, LS-R$^1_5$, LS-S$^1_4$, and LS-S$^1_5$ propagation schemes still allow for 
accelerating the calculations, although for some reactions the predicted energies are worse after propagation.

\begin{figure}[htbp]
      \centering
      \includegraphics[width=0.49\textwidth]{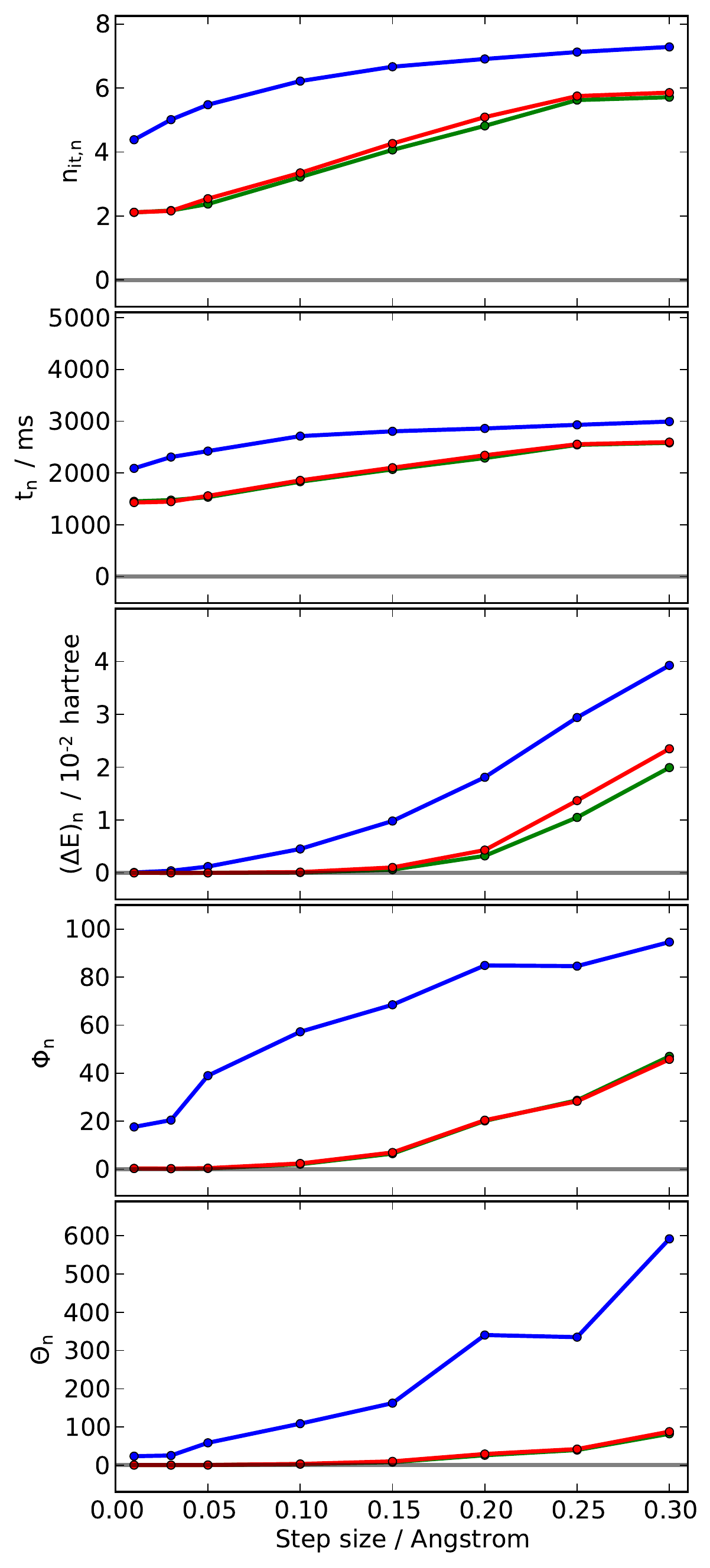}
      \caption{From top to bottom: Numbers of iterations $n_{\mathrm{it},n}$, execution times $t_n$, errors in the predicted energy $(\Delta E)_n$, in the gradient angle $\Phi_n$ and in the gradient amplitude $\Theta_n$ as specified in section \ref{sct:dr} for varying step sizes in PM6 calculations on the S$_\mathrm{N}$2 reaction (\textbf{B} in Fig.~\ref{reactions}).
      The color coding of the curves is identical to Fig.~\ref{comparison}.
      }
      \label{stepsize_picture}
\end{figure}

Fig.\ \ref{stepsize_picture} shows the efficiency of the LS-R$^1_4$ and LS-S$^1_4$ propagation schemes depending on the step size for the S$_\mathrm{N}$2 reaction with PM6.
With increasing step sizes, the SCF acceleration decreases, but always allow for faster calculations than when employing 
simply the last converged density matrix in combination with the DIIS algorithm (i.e., without propagation).

\section{Conclusions}

SCF methods in the real-time quantum chemistry framework introduce severe challenges due to their iterative nature.
To allow for more frequent feedback, one can attempt to accelerate SCF calculations. 
Various SCF acceleration schemes have already been proposed and are widely employed in quantum chemical calculations.
Many techniques reduce the computational cost of SCF calculations by combining results obtained for similar molecular structures.
Some of them rely on the continuity of the visited molecular structures in a dynamical sense and are 
therefore too restrictive for application in real-time quantum chemistry.

In this work, we improved SCF acceleration for series of molecular structures with moderate structural changes that lack time information and consistency in a dynamic sense, as is the case in real-time quantum chemistry.
To achieve this, our focus was on providing a good guess for the density matrix required for the first iteration of the SCF procedure 
in addition to the subsequent convergence acceleration, for which we chose the DIIS algorithm.

We developed two schemes that provide reliable guesses for the initial density matrix to reduce the number of SCF iterations
required for convergence.
The initial density matrix guess for the orbital optimization at a new structure was obtained from a linear extrapolation
of converged density matrices of structures visited before in the real-time exploration.
The coefficients of the linear extrapolation were determined in a least-squares minimization, 
analogously to the extrapolation of molecular orbitals by Atsumi and Nakai.\cite{atsumi2010}
We then allowed for partly restoring the idempotency of the density matrix guess by applying the McWeeny purification algorithm.
We denoted the schemes developed LS-R and LS-S.

Their application, combined with the DIIS acceleration, improved SCF convergence significantly for sequences of related molecular structures.
Within the real-time quantum chemistry framework, this allows for more frequent molecular-property updates, 
which improves the real-time exploration of reactivity and allows to study larger systems.
The developed schemes consistently improve SCF acceleration with respect to the application of the DIIS algorithm.
The magnitude of the acceleration depends on the similarity of the molecular structures and will be the larger, the 
smaller the structural changes between consecutive structures are.
For a set of six model reactions, we achieved a reduction in the number of SCF iterations of about 30--50\% for changes 
in atomic positions of up to 0.1~\AA{} between successive structures, which is commonly realized in real-time explorations
(note that the 0.1~\AA{} shift refers to individual nuclear coordinates rather than to a global shift of all coordinates).
For the PM6 and DFTB3 methods, this translates into a speedup of about 20--30\%.

Evaluating the energy and gradients with the propagated density matrix without SCF optimization and comparing them to the converged values is a good indicator of the accuracy of the propagated matrix.
Application of the LS-R and LS-S schemes consistently improved the predicted energies and gradients.
Subjecting the propagated density matrix to additional McWeeny purification steps delivered accurate predictions for this matrix.

Even though LS-R and LS-S were developed for application within real-time quantum chemistry, they will also be useful
for other approaches involving SCF iterations for sequences of related molecular structures.

\section*{Acknowledgments}
The authors gratefully acknowledge support by ETH Zurich (grant number: ETH-20 15-1).


\begin{thebibliography}{10}

\bibitem{haag2013}
Haag, M.~P.; Reiher, M. \emph{Int. J. Quantum Chem.} \textbf{2013}, \emph{113},
  8--20.

\bibitem{haag2014a}
Haag, M.~P.; Reiher, M. \emph{Faraday Discuss.} \textbf{2014}, \emph{169},
  89--118.

\bibitem{haag2014b}
Haag, M.~P.; Vaucher, A.~C.; Bosson, M.; Redon, S.; Reiher, M.
  \emph{ChemPhysChem} \textbf{2014}, \emph{15}, 3301--3319.

\bibitem{marti2009}
Marti, K.~H.; Reiher, M. \emph{J. Comput. Chem.} \textbf{2009}, \emph{30},
  2010--2020.

\bibitem{haag2011}
Haag, M.~P.; Marti, K.~H.; Reiher, M. \emph{ChemPhysChem} \textbf{2011},
  \emph{12}, 3204--3213.

\bibitem{vaucher2015a}
{Vaucher}, A.~C.; {Haag}, M.~P.; {Reiher}, M. \emph{ArXiv e-prints}
  \textbf{2015}.

\bibitem{luehr2015a}
Luehr, N.; Jin, A. G.~B.; Mart\'{i}nez, T.~J. \emph{J. Chem. Theory Comput.}
  \textbf{2015}, \emph{11}, 4536--4544.

\bibitem{gaus2011}
Gaus, M.; Cui, Q.; Elstner, M. \emph{J. Chem. Theory Comput.} \textbf{2011},
  \emph{7}, 931--948.

\bibitem{stewart2007}
Stewart, J. J.~P. \emph{J. Mol. Model.} \textbf{2007}, \emph{13}, 1173--1213.

\bibitem{korth2011c}
Korth, M.; Thiel, W. \emph{J. Chem. Theory Comput.} \textbf{2011}, \emph{7},
  2929--2936.

\bibitem{sedlak2013a}
Sedlak, R.; Janowski, T.; Pito\v{n}\'{a}k, M.; \v{R}ez\'{a}\v{c}, J.; Pulay,
  P.; Hobza, P. \emph{J. Chem. Theory Comput.} \textbf{2013}, \emph{9},
  3364--3374.

\bibitem{hostas2013a}
Hosta\v{s}, J.; \v{R}ez\'{a}\v{c}, J.; Hobza, P. \emph{Chem. Phys. Lett.}
  \textbf{2013}, \emph{568-569}, 161--166.

\bibitem{bosson2012}
Bosson, M.; Richard, C.; Plet, A.; Grudinin, S.; Redon, S. \emph{J. Comput.
  Chem.} \textbf{2012}, \emph{33}, 779--790.

\bibitem{elstner1998}
Elstner, M.; Porezag, D.; Jungnickel, G.; Elsner, J.; Haugk, M.; Frauenheim,
  T.; Suhai, S.; Seifert, G. \emph{Phys. Rev. B} \textbf{1998}, \emph{58},
  7260--7268.

\bibitem{saunders1973}
Saunders, V.~R.; Hillier, I.~H. \emph{Int. J. Quantum Chem.} \textbf{1973},
  \emph{7}, 699--705.

\bibitem{pulay1980}
Pulay, P. \emph{Chem. Phys. Lett.} \textbf{1980}, \emph{73}, 393--398.

\bibitem{pulay1982}
Pulay, P. \emph{J. Comput. Chem.} \textbf{1982}, \emph{3}, 556--560.

\bibitem{kudin2002}
Kudin, K.~N.; Scuseria, G.~E.; Cances, E. \emph{J. Chem. Phys.} \textbf{2002},
  \emph{116}, 8255--8261.

\bibitem{hu2010}
Hu, X.; Yang, W. \emph{J. Chem. Phys.} \textbf{2010}, \emph{132}, 054109.

\bibitem{host2008}
H{\o}st, S.; Olsen, J.; Jansik, B.; Th{\o}gersen, L.; J{\o}rgensen, P.;
  Helgaker, T. \emph{J. Chem. Phys.} \textbf{2008}, \emph{129}, 124106.

\bibitem{wang2011}
Wang, Y.~A.; Yam, C.~Y.; Chen, Y.~K.; Chen, G. \emph{J. Chem. Phys.}
  \textbf{2011}, \emph{134}, 241103.

\bibitem{chen2011a}
Chen, Y.~K.; Wang, Y.~A. \emph{J. Chem. Theory Comput.} \textbf{2011},
  \emph{7}, 3045--3048.

\bibitem{vandevondele2003}
VandeVondele, J.; Hutter, J. \emph{J. Chem. Phys.} \textbf{2003}, \emph{118},
  4365--4369.

\bibitem{car1985}
Car, R.; Parrinello, M. \emph{Phys. Rev. Lett.} \textbf{1985}, \emph{55},
  2471--2474.

\bibitem{schlegel2001}
Schlegel, H.~B.; Millam, J.~M.; Iyengar, S.~S.; Voth, G.~A.; Daniels, A.~D.;
  Scuseria, G.~E.; Frisch, M.~J. \emph{J. Chem. Phys.} \textbf{2001},
  \emph{114}, 9758--9763.

\bibitem{iyengar2001}
Iyengar, S.~S.; Schlegel, H.~B.; Millam, J.~M.; A.~Voth, G.; Scuseria, G.~E.;
  Frisch, M.~J. \emph{J. Chem. Phys.} \textbf{2001}, \emph{115}, 10291--10302.

\bibitem{schlegel2002a}
Schlegel, H.~B.; Iyengar, S.~S.; Li, X.; Millam, J.~M.; Voth, G.~A.; Scuseria,
  G.~E.; Frisch, M.~J. \emph{J. Chem. Phys.} \textbf{2002}, \emph{117},
  8694--8704.

\bibitem{niklasson2012}
Niklasson, A. M.~N.; Cawkwell, M.~J. \emph{Phys. Rev. B} \textbf{2012},
  \emph{86}, 174308.

\bibitem{souvatzis2014a}
Souvatzis, P.; Niklasson, A. M.~N. \emph{J. Chem. Phys.} \textbf{2014},
  \emph{140}, 044117.

\bibitem{pulay2004}
Pulay, P.; Fogarasi, G. \emph{Chem. Phys. Lett.} \textbf{2004}, \emph{386},
  272--278.

\bibitem{atsumi2008}
Atsumi, T.; Nakai, H. \emph{J. Chem. Phys.} \textbf{2008}, \emph{128}, 094101.

\bibitem{atsumi2010}
Atsumi, T.; Nakai, H. \emph{Chem. Phys. Lett.} \textbf{2010}, \emph{490},
  102--108.

\bibitem{vandevondele2005}
VandeVondele, J.; Krack, M.; Mohamed, F.; Parrinello, M.; Chassaing, T.;
  Hutter, J. \emph{Comput. Phys. Commun.} \textbf{2005}, \emph{167}, 103--128.

\bibitem{mcweeny1959}
McWeeny, R. \emph{Phys. Rev.} \textbf{1959}, \emph{114}, 1528--1529.

\bibitem{mcweeny1960}
McWeeny, R. \emph{Rev. Mod. Phys.} \textbf{1960}, \emph{32}, 335--369.

\end{thebibliography}

\providecommand{\url}[1]{\texttt{#1}}
\providecommand{\urlprefix}{}
\providecommand{\foreignlanguage}[2]{#2}
\providecommand{\Capitalize}[1]{\uppercase{#1}}
\providecommand{\capitalize}[1]{\expandafter\Capitalize#1}
\providecommand{\bibliographycite}[1]{\cite{#1}}
\providecommand{\bbland}{and}
\providecommand{\bblchap}{chap.}
\providecommand{\bblchapter}{chapter}
\providecommand{\bbletal}{et~al.}
\providecommand{\bbleditors}{editors}
\providecommand{\bbleds}{eds.}
\providecommand{\bbleditor}{editor}
\providecommand{\bbled}{ed.}
\providecommand{\bbledition}{edition}
\providecommand{\bbledn}{ed.}
\providecommand{\bbleidp}{page}
\providecommand{\bbleidpp}{pages}
\providecommand{\bblerratum}{erratum}
\providecommand{\bblin}{in}
\providecommand{\bblmthesis}{Master's thesis}
\providecommand{\bblno}{no.}
\providecommand{\bblnumber}{number}
\providecommand{\bblof}{of}
\providecommand{\bblpage}{page}
\providecommand{\bblpages}{pages}
\providecommand{\bblp}{p}
\providecommand{\bblphdthesis}{Ph.D. thesis}
\providecommand{\bblpp}{pp}
\providecommand{\bbltechrep}{Tech. Rep.}
\providecommand{\bbltechreport}{Technical Report}
\providecommand{\bblvolume}{volume}
\providecommand{\bblvol}{Vol.}
\providecommand{\bbljan}{January}
\providecommand{\bblfeb}{February}
\providecommand{\bblmar}{March}
\providecommand{\bblapr}{April}
\providecommand{\bblmay}{May}
\providecommand{\bbljun}{June}
\providecommand{\bbljul}{July}
\providecommand{\bblaug}{August}
\providecommand{\bblsep}{September}
\providecommand{\bbloct}{October}
\providecommand{\bblnov}{November}
\providecommand{\bbldec}{December}
\providecommand{\bblfirst}{First}
\providecommand{\bblfirsto}{1st}
\providecommand{\bblsecond}{Second}
\providecommand{\bblsecondo}{2nd}
\providecommand{\bblthird}{Third}
\providecommand{\bblthirdo}{3rd}
\providecommand{\bblfourth}{Fourth}
\providecommand{\bblfourtho}{4th}
\providecommand{\bblfifth}{Fifth}
\providecommand{\bblfiftho}{5th}
\providecommand{\bblst}{st}
\providecommand{\bblnd}{nd}
\providecommand{\bblrd}{rd}
\providecommand{\bblth}{th}

\clearpage
\newpage
\clearpage
\newpage
\begin{center}
\subsection*{Supporting Information}

{\large ``Accelerating Wave Function Convergence in Interactive Quantum Chemical Reactivity Studies''\\[1ex]
Adrian H.\ M\"{u}hlbach, Alain C.\ Vaucher, and Markus Reiher}\\[3ex]
{\small \textit{ETH Z\"urich, Laboratorium f\"ur Physikalische Chemie, Vladimir--Prelog-Weg 2, CH-8093 Z\"urich, Switzerland}}
\end{center}

%

\section*{Tables of Results}\label{tables}

The best performing propagation parameter combinations for each model reaction are highlighted in gray. 
Unless noted otherwise, values in gray for the number of iterations, execution times, energy error, gradient angle and amplitude error perform within the best 10\%, 5\%, 50\%, 25\% and 50\%, respectively. 
In every table, the last row (denoted by \o) presents results from calculations where no propagation was applied.
The equations specified in the table captions refer to the main text of the article.

The execution times were obtained by averaging the timings of ten runs on four threads on a machine with a 3.40GHz Intel Xeon E3-1240 v2 870 CPU. 

  \newpage

    \begin{table}[hb]
    \centering\small
\begin{tabular}{cccccccccc} 
    \toprule
    \multicolumn{1}{c}{$n_{it}$} & \multicolumn{6}{c}{PM6} & \multicolumn{3}{c}{DFTB3}\\
    \cmidrule(lr){2-7}\cmidrule(lr){8-10}
      & \textbf{A} & \textbf{B} & \textbf{C} & \textbf{D} & \textbf{E} & \textbf{F} & \textbf{A} & \textbf{E} & \textbf{F}\\ 
    \midrule
LS-R$_3^0$ & 3.7 & 4.2 & 5.9 & 6.7 & 3.8 & 5.1 & 3.9 & \ccell[gray]{0.9}{5.1} & \ccell[gray]{0.9}{3.6} \\
LS-R$_3^1$ & \ccell[gray]{0.9}{3.6} & 4.0 & 5.6 & \ccell[gray]{0.9}{6.4} & 3.6 & \ccell[gray]{0.9}{4.5} & 3.6 & \ccell[gray]{0.9}{4.8} & \ccell[gray]{0.9}{3.4} \\
LS-R$_3^2$ & \ccell[gray]{0.9}{3.6} & 4.0 & 5.6 & \ccell[gray]{0.9}{6.3} & 3.6 & 5.5 & \ccell[gray]{0.9}{3.6} & \ccell[gray]{0.9}{4.6} & \ccell[gray]{0.9}{3.4} \\
LS-R$_3^3$ & \ccell[gray]{0.9}{3.6} & 4.0 & 5.6 & \ccell[gray]{0.9}{6.3} & 3.6 & 5.1 & \ccell[gray]{0.9}{3.6} & \ccell[gray]{0.9}{4.6} & \ccell[gray]{0.9}{3.4} \\
LS-R$_3^4$ & \ccell[gray]{0.9}{3.6} & 4.0 & 5.6 & \ccell[gray]{0.9}{6.3} & 3.6 & 5.4 & \ccell[gray]{0.9}{3.6} & \ccell[gray]{0.9}{4.6} & \ccell[gray]{0.9}{3.4} \\
    \midrule
LS-R$_4^0$ & \ccell[gray]{0.9}{3.5} & \ccell[gray]{0.9}{3.6} & 4.8 & 6.4 & \ccell[gray]{0.9}{3.4} & 6.3 & 3.6 & 5.4 & \ccell[gray]{0.9}{3.6} \\
\bottomrule
    \end{tabular}
    \caption{Number of SCF iterations $n_{\mathrm{it}}$ of Eq.\ (\ref{dr:iterations}) needed to reach convergence after employing the LS-R propagation scheme.
             The values are averages over the different molecular structures along the reaction coordinates.}
    \label{lsdmr-it}
    \end{table}

    \begin{table}[hp]
    \centering\small
    \begin{tabular}{cccccccccc} 
    \toprule
    \multicolumn{1}{c}{$t$} & \multicolumn{6}{c}{PM6} & \multicolumn{3}{c}{DFTB3}\\
    \cmidrule(lr){2-7}\cmidrule(lr){8-10}
      & \textbf{A} & \textbf{B} & \textbf{C} & \textbf{D} & \textbf{E} & \textbf{F} & \textbf{A} & \textbf{E} & \textbf{F}\\ 
    \midrule
LS-R$_3^0$ & \ccell[gray]{0.9}{1.17} & 2.02 & 5.36 & 40.07 & 3.96 & 3.81 & 1.01 & \ccell[gray]{0.9}{8.86} & 2.76 \\
LS-R$_3^1$ & \ccell[gray]{0.9}{1.17} & 2.02 & 5.22 & \ccell[gray]{0.9}{38.96} & 3.95 & \ccell[gray]{0.9}{3.44} & 0.98 & \ccell[gray]{0.9}{8.64} & \ccell[gray]{0.9}{2.68} \\
LS-R$_3^2$ & 1.19 & 2.05 & 5.30 & \ccell[gray]{0.9}{39.02} & 4.04 & 3.78 & 1.00 & \ccell[gray]{0.9}{8.66} & 2.85 \\
LS-R$_3^3$ & 1.21 & 2.08 & 5.37 & 39.34 & 4.13 & 3.88 & 1.03 & \ccell[gray]{0.9}{8.91} & 2.94 \\
LS-R$_3^4$ & 1.23 & 2.11 & 5.45 & 39.68 & 4.22 & 3.95 & 1.06 & 9.16 & 3.03 \\
    \midrule
LS-R$_4^0$ & \ccell[gray]{0.9}{1.13} & \ccell[gray]{0.9}{1.86} & 4.68 & \ccell[gray]{0.9}{39.03} & \ccell[gray]{0.9}{3.68} & 4.13 & \ccell[gray]{0.9}{0.96} & 9.48 & 2.74 \\
\bottomrule
    \end{tabular}
    \caption{Computational time $t$ (in ms) of Eq.\ (\ref{dr:speed}) when employing the LS-R propagation scheme.}
    \label{lsdmr-t}
    \end{table}

    \begin{table}[hp]
    \centering\small
    \begin{tabular}{cccccccccc} 
    \toprule
    \multicolumn{1}{c}{$\Delta E$} & \multicolumn{6}{c}{PM6} & \multicolumn{3}{c}{DFTB3}\\
    \cmidrule(lr){2-7}\cmidrule(lr){8-10}
      & \textbf{A} & \textbf{B} & \textbf{C} & \textbf{D} & \textbf{E} & \textbf{F} & \textbf{A} & \textbf{E} & \textbf{F}\\ 
    \midrule
LS-R$_3^0$ & 1210.1 & 773.1 & 794.3 & 3299.6 & 132.9 & 136.3 & 183.8 & 363.1 & 93.7 \\
LS-R$_3^1$ & 268.9 & 44.5 & 18.0 & 1184.8 & \ccell[gray]{0.9}{1.1} & \ccell[gray]{0.9}{1.1} & 11.2 & 7.5 & 2.6 \\
LS-R$_3^2$ & 40.5 & 8.0 & \ccell[gray]{0.9}{6.6} & 506.0 & \ccell[gray]{0.9}{1.1} & \ccell[gray]{0.9}{0.9} & 1.7 & \ccell[gray]{0.9}{2.0} & \ccell[gray]{0.9}{1.1} \\
LS-R$_3^3$ & 19.1 & 7.8 & \ccell[gray]{0.9}{6.6} & 176.6 & \ccell[gray]{0.9}{1.1} & \ccell[gray]{0.9}{1.2} & 1.7 & \ccell[gray]{0.9}{2.0} & \ccell[gray]{0.9}{1.1} \\
LS-R$_3^4$ & 18.9 & 7.8 & \ccell[gray]{0.9}{6.6} & \ccell[gray]{0.9}{89.1} & \ccell[gray]{0.9}{1.1} & \ccell[gray]{0.9}{0.8} & 1.7 & \ccell[gray]{0.9}{2.0} & \ccell[gray]{0.9}{1.1} \\
    \midrule
LS-R$_4^0$ & 804.5 & 410.2 & 439.1 & 2698.5 & 35.4 & 73.2 & 88.0 & 565.3 & 50.6 \\
\bottomrule
    \end{tabular}
    \caption{Error in the energy $\Delta E$ (in $10^{-5}$ Hartree) of Eq.\ (\ref{dr:energy}) after employing the LS-R propagation scheme.}
    \label{lsdmr-e}
    \end{table}

    \begin{table}[hp]
    \centering\small
    \begin{tabular}{cccccccccc} 
    \toprule
    \multicolumn{1}{c}{$\Phi$} & \multicolumn{6}{c}{PM6} & \multicolumn{3}{c}{DFTB3}\\
    \cmidrule(lr){2-7}\cmidrule(lr){8-10}
      & \textbf{A} & \textbf{B} & \textbf{C} & \textbf{D} & \textbf{E} & \textbf{F} & \textbf{A} & \textbf{E} & \textbf{F}\\ 
    \midrule
LS-R$_3^0$ & 3.6 & 4.6 & 17.6 & 11.0 & 4.0 & 2.9 & 1.9 & \ccell[gray]{0.9}{8.6} & 1.5 \\
LS-R$_3^1$ & 3.0 & 3.7 & 15.5 & 10.3 & 3.7 & \ccell[gray]{0.9}{1.1} & 1.6 & \ccell[gray]{0.9}{9.3} & 1.3 \\
LS-R$_3^2$ & 3.0 & 3.7 & 15.5 & 10.0 & 3.7 & \ccell[gray]{0.9}{1.1} & 1.6 & \ccell[gray]{0.9}{9.2} & 1.3 \\
LS-R$_3^3$ & 3.0 & 3.7 & 15.5 & 10.0 & 3.7 & 1.9 & 1.6 & \ccell[gray]{0.9}{9.2} & 1.3 \\
LS-R$_3^4$ & 3.0 & 3.7 & 15.5 & 10.1 & 3.7 & 1.3 & 1.6 & \ccell[gray]{0.9}{9.2} & 1.3 \\
    \midrule
LS-R$_4^0$ & 2.9 & 2.9 & 7.8 & 9.7 & 1.7 & 2.0 & 1.1 & \ccell[gray]{0.9}{8.9} & 1.1 \\
\bottomrule
    \end{tabular}
    \caption{Gradient angle error $\Phi$ (in degrees) of Eq.\ (\ref{dr:angle}) after employing the LS-R propagation scheme.}
    \label{lsdmr-ang}
    \end{table}

    \begin{table}[hp]
    \centering\small
    \begin{tabular}{cccccccccc} 
    \toprule
    \multicolumn{1}{c}{$\Theta$} & \multicolumn{6}{c}{PM6} & \multicolumn{3}{c}{DFTB3}\\
    \cmidrule(lr){2-7}\cmidrule(lr){8-10}
      & \textbf{A} & \textbf{B} & \textbf{C} & \textbf{D} & \textbf{E} & \textbf{F} & \textbf{A} & \textbf{E} & \textbf{F}\\ 
    \midrule
LS-R$_3^0$ & 16.5 & 6.9 & 23.8 & 33.6 & 8.4 & 2.7 & 2.5 & \ccell[gray]{0.9}{22.9} & 17.5 \\
LS-R$_3^1$ & 9.1 & 5.3 & 22.7 & 25.3 & 7.5 & \ccell[gray]{0.9}{2.2} & 2.1 & \ccell[gray]{0.9}{19.9} & 25.3 \\
LS-R$_3^2$ & 10.1 & 5.3 & 22.8 & 25.7 & 7.5 & 2.6 & 2.2 & \ccell[gray]{0.9}{19.9} & 25.0 \\
LS-R$_3^3$ & 10.2 & 5.3 & 22.8 & 26.3 & 7.5 & 2.9 & 2.2 & \ccell[gray]{0.9}{19.9} & 25.0 \\
LS-R$_3^4$ & 10.2 & 5.3 & 22.8 & 26.5 & 7.5 & 2.9 & 2.2 & \ccell[gray]{0.9}{19.9} & 25.0 \\
    \midrule
LS-R$_4^0$ & 11.0 & \ccell[gray]{0.9}{3.7} & 12.8 & 28.2 & \ccell[gray]{0.9}{2.2} & 2.5 & \ccell[gray]{0.9}{1.4} & 28.3 & 2.1 \\
\bottomrule
    \end{tabular}
    \caption{Gradient amplitude error $\Theta$ (in percent) of Eq.\ (\ref{dr:amplitude}) after employing the LS-R propagation scheme.}
    \label{lsdmr-amp}
    \end{table}

\clearpage
  \newpage

    \begin{table}[hb]
    \centering\small
    \begin{tabular}{cccccccccc} 
    \toprule
    \multicolumn{1}{c}{$n_{it}$} & \multicolumn{6}{c}{PM6} & \multicolumn{3}{c}{DFTB3}\\
    \cmidrule(lr){2-7}\cmidrule(lr){8-10}
      & \textbf{A} & \textbf{B} & \textbf{C} & \textbf{D} & \textbf{E} & \textbf{F} & \textbf{A} & \textbf{E} & \textbf{F}\\ 
    \midrule
LS-S$_3^0$ & 3.8 & 4.2 & 5.9 & 6.4 & 3.7 & \ccell[gray]{0.9}{4.7} & 3.6 & 4.9 & 3.4 \\
LS-S$_3^1$ & \ccell[gray]{0.9}{3.6} & 3.9 & 5.6 & 6.2 & 3.6 & \ccell[gray]{0.9}{4.3} & \ccell[gray]{0.9}{3.5} & \ccell[gray]{0.9}{4.6} & \ccell[gray]{0.9}{3.0} \\
LS-S$_3^2$ & \ccell[gray]{0.9}{3.5} & 3.9 & 5.6 & 6.2 & 3.6 & 5.2 & \ccell[gray]{0.9}{3.5} & \ccell[gray]{0.9}{4.6} & \ccell[gray]{0.9}{3.1} \\
LS-S$_3^3$ & \ccell[gray]{0.9}{3.5} & 3.9 & 5.6 & \ccell[gray]{0.9}{6.1} & 3.6 & 5.0 & \ccell[gray]{0.9}{3.5} & \ccell[gray]{0.9}{4.6} & \ccell[gray]{0.9}{3.1} \\
LS-S$_3^4$ & \ccell[gray]{0.9}{3.5} & 3.9 & 5.6 & \ccell[gray]{0.9}{6.1} & 3.6 & 5.2 & \ccell[gray]{0.9}{3.5} & \ccell[gray]{0.9}{4.6} & \ccell[gray]{0.9}{3.1} \\
    \midrule
LS-S$_4^0$ & \ccell[gray]{0.9}{3.4} & \ccell[gray]{0.9}{3.5} & 4.8 & 6.2 & 3.4 & 5.5 & 3.6 & 4.9 & \ccell[gray]{0.9}{3.3} \\
\bottomrule
    \end{tabular}
    \caption{Number of SCF iterations $n_\mathrm{it}$ of Eq.\ (\ref{dr:iterations}) needed to reach convergence after employing the LS-S propagation scheme.
             The values are averages over the different molecular structures along the reaction coordinates.}
    \label{lsdms-it}
    \end{table}

    \begin{table}[hp]
    \centering\small
    \begin{tabular}{cccccccccc} 
    \toprule
    \multicolumn{1}{c}{$t$} & \multicolumn{6}{c}{PM6} & \multicolumn{3}{c}{DFTB3}\\
    \cmidrule(lr){2-7}\cmidrule(lr){8-10}
      & \textbf{A} & \textbf{B} & \textbf{C} & \textbf{D} & \textbf{E} & \textbf{F} & \textbf{A} & \textbf{E} & \textbf{F}\\ 
    \midrule
LS-S$_3^0$ & \ccell[gray]{0.9}{1.19} & 2.03 & 5.38 & 39.14 & 3.95 & \ccell[gray]{0.9}{3.34} & \ccell[gray]{0.9}{0.98} & \ccell[gray]{0.9}{8.61} & 2.65 \\
LS-S$_3^1$ & \ccell[gray]{0.9}{1.18} & 1.99 & 5.25 & \ccell[gray]{0.9}{38.53} & 3.94 & \ccell[gray]{0.9}{3.41} & \ccell[gray]{0.9}{0.98} & \ccell[gray]{0.9}{8.39} & \ccell[gray]{0.9}{2.43} \\
LS-S$_3^2$ & \ccell[gray]{0.9}{1.19} & 2.02 & 5.39 & \ccell[gray]{0.9}{38.82} & 4.03 & \ccell[gray]{0.9}{3.44} & \ccell[gray]{0.9}{1.00} & \ccell[gray]{0.9}{8.53} & 2.60 \\
LS-S$_3^3$ & 1.21 & 2.05 & 5.39 & 38.93 & 4.12 & 3.88 & 1.03 & 8.78 & 2.69 \\
LS-S$_3^4$ & 1.23 & 2.08 & 5.47 & 39.25 & 4.21 & 3.79 & 1.06 & 9.03 & 2.78 \\
    \midrule
LS-S$_4^0$ & \ccell[gray]{0.9}{1.14} & \ccell[gray]{0.9}{1.85} & 4.67 & \ccell[gray]{0.9}{38.60} & 3.80 & 3.50 & \ccell[gray]{0.9}{1.00} & 8.72 & 2.57 \\
\bottomrule
    \end{tabular}
    \caption{Computational time $t$ (in ms) Eq.\ (\ref{dr:speed}) when employing the LS-S propagation scheme.}
    \label{lsdms-t}
    \end{table}

    \begin{table}[hp]
    \centering\small
    \begin{tabular}{cccccccccc} 
    \toprule
    \multicolumn{1}{c}{$\Delta E$} & \multicolumn{6}{c}{PM6} & \multicolumn{3}{c}{DFTB3}\\
    \cmidrule(lr){2-7}\cmidrule(lr){8-10}
      & \textbf{A} & \textbf{B} & \textbf{C} & \textbf{D} & \textbf{E} & \textbf{F} & \textbf{A} & \textbf{E} & \textbf{F}\\ 
    \midrule
LS-S$_3^0$ & 1086.4 & 503.4 & 734.5 & 2216.7 & 82.7 & 60.6 & 209.4 & 264.9 & 21.5 \\
LS-S$_3^1$ & 229.1 & 22.6 & 18.6 & 615.4 & \ccell[gray]{0.9}{1.0} & \ccell[gray]{0.9}{0.9} & 11.1 & 3.6 & \ccell[gray]{0.9}{1.1} \\
LS-S$_3^2$ & 29.1 & 5.8 & \ccell[gray]{0.9}{7.0} & 193.0 & \ccell[gray]{0.9}{1.0} & \ccell[gray]{0.9}{0.9} & 1.6 & \ccell[gray]{0.9}{1.3} & \ccell[gray]{0.9}{0.8} \\
LS-S$_3^3$ & 12.4 & 5.8 & \ccell[gray]{0.9}{7.0} & 74.5 & \ccell[gray]{0.9}{1.0} & \ccell[gray]{0.9}{0.9} & 1.6 & \ccell[gray]{0.9}{1.3} & \ccell[gray]{0.9}{0.8} \\
LS-S$_3^4$ & 12.3 & 5.8 & \ccell[gray]{0.9}{7.0} & 62.9 & \ccell[gray]{0.9}{1.0} & \ccell[gray]{0.9}{0.9} & 1.6 & \ccell[gray]{0.9}{1.3} & \ccell[gray]{0.9}{0.8} \\
    \midrule
LS-S$_4^0$ & 642.2 & 208.0 & 368.4 & 1978.7 & 22.3 & 35.0 & 98.9 & 231.7 & 13.7 \\
\bottomrule
    \end{tabular}
    \caption{Error in the energy $\Delta E$ (in $10^{-5}$ Hartree) of Eq.\ (\ref{dr:energy}) after employing the LS-S propagation scheme.}
    \label{lsdms-e}
    \end{table}

    \begin{table}[hp]
    \centering\small
    \begin{tabular}{cccccccccc} 
    \toprule
    \multicolumn{1}{c}{$\Phi$} & \multicolumn{6}{c}{PM6} & \multicolumn{3}{c}{DFTB3}\\
    \cmidrule(lr){2-7}\cmidrule(lr){8-10}
      & \textbf{A} & \textbf{B} & \textbf{C} & \textbf{D} & \textbf{E} & \textbf{F} & \textbf{A} & \textbf{E} & \textbf{F}\\ 
    \midrule
LS-S$_3^0$ & 3.3 & 4.2 & 18.2 & 10.9 & 3.9 & 1.3 & 1.9 & 7.8 & 1.1 \\
LS-S$_3^1$ & 2.7 & 3.5 & 16.3 & 10.4 & 3.6 & 1.1 & 1.8 & 8.2 & 0.9 \\
LS-S$_3^2$ & 2.8 & 3.5 & 16.3 & 10.2 & 3.6 & \ccell[gray]{0.9}{0.8} & 1.7 & 8.2 & 0.9 \\
LS-S$_3^3$ & 2.8 & 3.5 & 16.3 & 10.3 & 3.6 & \ccell[gray]{0.9}{0.8} & 1.7 & 8.2 & 0.9 \\
LS-S$_3^4$ & 2.8 & 3.5 & 16.3 & 10.3 & 3.6 & \ccell[gray]{0.9}{0.8} & 1.7 & 8.2 & 0.9 \\
    \midrule
LS-S$_4^0$ & 2.3 & 2.5 & 8.0 & 9.0 & 1.5 & 3.9 & 1.1 & \ccell[gray]{0.9}{7.2} & \ccell[gray]{0.9}{0.8} \\
\bottomrule
    \end{tabular}
    \caption{Gradient angle error $\Phi$ (in degrees) Eq.\ (\ref{dr:angle}) after employing the LS-S propagation scheme.}
    \label{lsdms-ang}
    \end{table}

    \begin{table}[hp]
    \centering\small
    \begin{tabular}{cccccccccc} 
    \toprule
    \multicolumn{1}{c}{$\Theta$} & \multicolumn{6}{c}{PM6} & \multicolumn{3}{c}{DFTB3}\\
    \cmidrule(lr){2-7}\cmidrule(lr){8-10}
      & \textbf{A} & \textbf{B} & \textbf{C} & \textbf{D} & \textbf{E} & \textbf{F} & \textbf{A} & \textbf{E} & \textbf{F}\\ 
    \midrule
LS-S$_3^0$ & 12.7 & 6.2 & 25.2 & 27.6 & 6.1 & \ccell[gray]{0.9}{1.1} & 2.5 & 15.0 & 6.3 \\
LS-S$_3^1$ & 6.8 & 5.1 & 22.9 & 22.7 & 5.4 & \ccell[gray]{0.9}{1.1} & 2.3 & 14.2 & 9.0 \\
LS-S$_3^2$ & 7.3 & 5.1 & 23.0 & 23.2 & 5.4 & \ccell[gray]{0.9}{1.3} & 2.3 & 14.3 & 6.9 \\
LS-S$_3^3$ & 7.4 & 5.1 & 23.0 & 23.5 & 5.4 & \ccell[gray]{0.9}{1.1} & 2.3 & 14.3 & 6.9 \\
LS-S$_3^4$ & 7.4 & 5.1 & 23.0 & 23.5 & 5.4 & \ccell[gray]{0.9}{1.2} & 2.3 & 14.3 & 6.9 \\
    \midrule
LS-S$_4^0$ & 5.9 & \ccell[gray]{0.9}{3.1} & 12.6 & 22.7 & \ccell[gray]{0.9}{1.7} & 4.6 & \ccell[gray]{0.9}{1.2} & \ccell[gray]{0.9}{12.9} & \ccell[gray]{0.9}{0.8} \\
\bottomrule
    \end{tabular}
    \caption{Gradient amplitude error $\Theta$ (in percent) of Eq.\ (\ref{dr:amplitude}) after employing the LS-S propagation scheme.}
    \label{lsdms-amp}
    \end{table}

\clearpage
  \newpage

    \begin{table}[hp]
    \centering\small
    \begin{tabular}{cccccccccc} 
    \toprule
    \multicolumn{1}{c}{$n_{it}$} & \multicolumn{6}{c}{PM6} & \multicolumn{3}{c}{DFTB3}\\
    \cmidrule(lr){2-7}\cmidrule(lr){8-10}
      & \textbf{A} & \textbf{B} & \textbf{C} & \textbf{D} & \textbf{E} & \textbf{F} & \textbf{A} & \textbf{E} & \textbf{F}\\ 
    \midrule
LS-R$_4^1$ & 5.1 & 5.5 & 7.5 & 8.7 & 5.2 & 5.5 & 4.2 & 5.4 & 3.4 \\
LS-R$_5^1$ & 5.1 & 5.1 & 7.5 & 8.8 & 4.7 & 6.0 & 4.2 & 5.6 & 3.5 \\
    \midrule
LS-S$_4^1$ & 5.1 & 5.2 & 7.5 & 8.8 & 4.8 & 5.4 & 4.2 & 5.4 & 3.2 \\
LS-S$_5^1$ & 5.0 & 5.1 & 7.2 & 8.8 & 4.3 & 6.2 & 4.2 & 5.6 & 3.4 \\
    \midrule
    \midrule
\o & 6.8 & 7.2 & 9.5 & 9.3 & 8.3 & 11.0 & 5.2 & 7.2 & 4.5 \\
    \bottomrule
    \end{tabular}
    \caption{Number of SCF iterations $n_\mathrm{it}$ of Eq.\ (\ref{dr:iterations}) needed to reach convergence for model reactions with doubled step size.
             The values are averages over the different molecular structures along the reaction coordinates.}
    \label{comp-step-it}
    \end{table}

    \begin{table}[hp]
    \centering\small
    \begin{tabular}{cccccccccc} 
    \toprule
    \multicolumn{1}{c}{$t$} & \multicolumn{6}{c}{PM6} & \multicolumn{3}{c}{DFTB3}\\
    \cmidrule(lr){2-7}\cmidrule(lr){8-10}
      & \textbf{A} & \textbf{B} & \textbf{C} & \textbf{D} & \textbf{E} & \textbf{F} & \textbf{A} & \textbf{E} & \textbf{F}\\ 
    \midrule
LS-R$_4^1$ & 1.46 & 2.48 & 6.44 & 48.63 & 4.98 & 3.88 & 1.13 & 10.24 & 2.84 \\
LS-R$_5^1$ & 1.45 & 2.36 & 6.45 & 49.34 & 4.65 & 4.17 & 1.13 & 10.63 & 2.94 \\
    \midrule
LS-S$_4^1$ & 1.48 & 2.43 & 6.50 & 49.71 & 4.82 & 3.84 & 1.16 & 9.68 & 2.61 \\
LS-S$_5^1$ & 1.47 & 2.51 & 6.37 & 50.06 & 4.52 & 4.36 & 1.16 & 10.08 & 2.72 \\
    \midrule
    \midrule
\o & 1.74 & 2.97 & 7.63 & 50.69 & 6.96 & 6.93 & 1.29 & 12.24 & 3.32 \\
    \bottomrule
    \end{tabular}
    \caption{Computational time $t$ (in ms) of Eq.\ (\ref{dr:speed}) for model reactions with doubled step size. }
    \label{comp-step-t}
    \end{table}

    \begin{table}[hp]
    \centering\small
    \begin{tabular}{cccccccccc} 
    \toprule
    \multicolumn{1}{c}{$\Delta E$} & \multicolumn{6}{c}{PM6} & \multicolumn{3}{c}{DFTB3}\\
    \cmidrule(lr){2-7}\cmidrule(lr){8-10}
      & \textbf{A} & \textbf{B} & \textbf{C} & \textbf{D} & \textbf{E} & \textbf{F} & \textbf{A} & \textbf{E} & \textbf{F}\\ 
    \midrule
LS-R$_4^1$ & 2715.5 & 590.2 & 691.3 & 10345.8 & 11.3 & 3.5 & 218.6 & 233.9 & 12.7 \\
LS-R$_5^1$ & 2827.8 & 579.0 & 621.3 & 9990.7 & 2.9 & 0.8 & 192.3 & 143.6 & 17.0 \\
    \midrule
LS-S$_4^1$ & 2085.1 & 439.5 & 705.2 & 6280.4 & 6.8 & 0.6 & 249.9 & 15.1 & 2.7 \\
LS-S$_5^1$ & 2169.5 & 346.9 & 468.8 & 6305.7 & 2.3 & 0.9 & 196.4 & 55.0 & 2.4 \\
    \midrule
    \midrule
\o & 1661.4 & 2351.5 & 3759.6 & 5704.8 & 1263.0 & 60.6 & 2489.5 & 758.0 & 63.7 \\
    \bottomrule
    \end{tabular}
    \caption{Error in the energy $\Delta E$ (in $10^{-5}$ Hartree) of Eq.\ (\ref{dr:energy}) for model reactions with doubled step size. }
    \label{comp-step-e}
    \end{table}

    \begin{table}[hp]
    \centering\small
    \begin{tabular}{cccccccccc} 
    \toprule
    \multicolumn{1}{c}{$\Phi$} & \multicolumn{6}{c}{PM6} & \multicolumn{3}{c}{DFTB3}\\
    \cmidrule(lr){2-7}\cmidrule(lr){8-10}
      & \textbf{A} & \textbf{B} & \textbf{C} & \textbf{D} & \textbf{E} & \textbf{F} & \textbf{A} & \textbf{E} & \textbf{F}\\ 
    \midrule
LS-R$_4^1$ & 16.4 & 24.2 & 54.6 & 36.5 & 12.1 & 3.4 & 7.3 & 31.0 & 2.2 \\
LS-R$_5^1$ & 15.7 & 20.2 & 46.8 & 34.7 & 9.5 & 2.6 & 7.4 & 33.9 & 2.6 \\
    \midrule
LS-S$_4^1$ & 15.1 & 24.2 & 51.5 & 37.8 & 11.5 & 1.6 & 8.8 & 31.5 & 0.9 \\
LS-S$_5^1$ & 13.6 & 19.6 & 48.6 & 35.8 & 8.9 & 2.7 & 7.5 & 35.1 & 1.3 \\
    \midrule
    \midrule
\o & 66.8 & 94.1 & 93.6 & 73.4 & 51.1 & 10.3 & 43.3 & 50.3 & 4.7 \\
    \bottomrule
    \end{tabular}
    \caption{Gradient angle error $\Phi$ (in degrees) Eq.\ (\ref{dr:angle}) for model reactions with doubled step size.}
    \label{comp-step-ang}
    \end{table}

    \begin{table}[hp]
    \centering\small
    \begin{tabular}{cccccccccc} 
    \toprule
    \multicolumn{1}{c}{$\Theta$} & \multicolumn{6}{c}{PM6} & \multicolumn{3}{c}{DFTB3}\\
    \cmidrule(lr){2-7}\cmidrule(lr){8-10}
      & \textbf{A} & \textbf{B} & \textbf{C} & \textbf{D} & \textbf{E} & \textbf{F} & \textbf{A} & \textbf{E} & \textbf{F}\\ 
    \midrule
LS-R$_4^1$ & 62.3 & 33.7 & 124.9 & 248.9 & 46.4 & 3.7 & 18.0 & 150.7 & 2.0 \\
LS-R$_5^1$ & 59.2 & 30.5 & 106.4 & 240.9 & 26.2 & 3.0 & 17.0 & 118.9 & 3.3 \\
    \midrule
LS-S$_4^1$ & 51.4 & 29.8 & 138.0 & 221.9 & 34.3 & 1.1 & 20.6 & 84.7 & 1.5 \\
LS-S$_5^1$ & 47.5 & 26.3 & 109.4 & 220.1 & 19.3 & 3.1 & 17.6 & 91.2 & 1.6 \\
    \midrule
    \midrule
\o & 262.3 & 351.7 & 1769.9 & 495.6 & 1890.4 & 41.4 & 137.7 & 983.9 & 35.1 \\
    \bottomrule
    \end{tabular}
    \caption{Gradient amplitude error $\Theta$ (in percent) of Eq.\ (\ref{dr:amplitude}) for model reactions with doubled step size. }
    \label{comp-step-amp}
    \end{table}

\end{document}